\documentclass[journal]{IEEEtranTIE}
\usepackage{amsmath,amsfonts}
\usepackage{algorithmic}
\usepackage{array}
\usepackage[caption=false,font=footnotesize,labelfont=rm,textfont=sf]{subfig}
\usepackage{textcomp}
\usepackage{stfloats}
\usepackage{url}
\usepackage{verbatim}
\usepackage{graphicx}
\def\BibTeX{{\rm B\kern-.05em{\sc i\kern-.025em b}\kern-.08em
    T\kern-.1667em\lower.7ex\hbox{E}\kern-.125emX}}
\usepackage{balance}
\usepackage{flushend}           

\usepackage{amssymb} 
\usepackage{mathtools}
\usepackage{booktabs}
\usepackage{upgreek}
\usepackage{soul,xcolor}
\usepackage{empheq}
\allowdisplaybreaks 

\begin{document}

\title{\huge{On the Stability of Electromechanical Switching~Devices}}

\author{
	
	Edgar Ramirez-Laboreo, 
    Carlos Sagues, 
	Eduardo Moya-Lasheras,
	and Eloy Serrano-Seco

	\thanks{
	
        This work was supported in part via grants PID2021-124137OB-I00, TED2021-130224B-I00, and CPP2021-008938, funded by MCIN/AEI/10.13039/501100011033 and by ERDF A way of making Europe; in part by the Ministerio de Educaci\'on, Cultura y Deporte, Gobierno de Espa\~na (grant FPU14/04171); in part by the Government of Arag\'on (grant T45{\_}23R), and in part by the ``Programa Investigo'' funded by the European Union NextGenerationEU.

        The authors are with the Departamento de Informatica e Ingenieria de Sistemas (DIIS) and the Instituto de Investigacion en Ingenieria de Aragon (I3A), Universidad de Zaragoza, 50018 Zaragoza, Spain (e-mail: ramirlab@unizar.es; csagues@unizar.es; emoya@unizar.es; eserranoseco@unizar.es ). 
				
	}
}

\maketitle

\begin{abstract}
Electromagnetic relays and solenoid actuators are commonly used for their bistable behavior, which allows for switching between two states in electrical, pneumatic, or hydraulic circuits, among other applications. Although there has been extensive research on modeling, estimation, and control of these electromechanical systems, a gap remains in the analysis area. This paper addresses this gap by presenting an equilibrium and stability analysis to gain deeper insight into their bistability. This analysis leverages a hybrid dynamical model to obtain analytical expressions that relate the physical parameters to the switching conditions. These expressions are useful, e.g., for fundamental understanding, quick analyses, or design optimization. The results are discussed in depth and potential practical applications are explored. Finally, the analysis is validated with experimental results from a real device.
\end{abstract}


\section{Introduction}

\IEEEPARstart{E}{lectromechanical} switching devices like solenoid actuators, electromagnetic relays, and on/off valves are commercial devices that are extensively used in several present-day applications, e.g., battery chargers for electric vehicles \cite{haghbin2013grid}, electronic stability control systems \cite{Zhao2016linear}, home appliances {\cite{acero2010domestic}}, or soft robotics~{\cite{xavier2022}}, among others. These devices exhibit bistable behavior, i.e., they switch between two positions depending on the supply voltage. For this reason, they are very well suited for modifying the configuration of electrical, pneumatic, or hydraulic circuits, and also as low-power actuators in simple mechanisms.

\begin{figure}[t]
\centering
\includegraphics[]{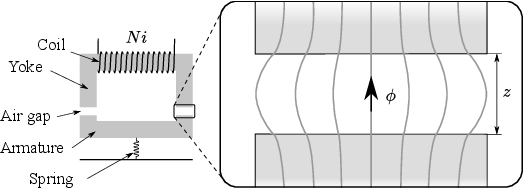}%
\caption{Diagram of a single-coil c-core reluctance actuator.}
\label{fig:MEC}
\end{figure}

These switching devices are all based on a small single-coil reluctance actuator with a limited range of motion. In essence, a reluctance actuator is an electromagnet, constructed by wrapping a wire around a ferromagnetic core, together with a movable component known as armature (see Fig.~\ref{fig:MEC}). When the coil is supplied with power, a magnetic force that pulls the armature toward the yoke is created. This force is always attractive, so the reverse motion is usually produced by a spring or simply by gravity. More advanced actuators can be designed from this concept, e.g., by including permanent magnets to reduce power consumption \cite{guofu2012permanent,huang2022modeling} or a second coil to increase control possibilities \cite{ito2019flux,su2022neural}.

Several research works have been devoted to the modeling and control of these devices. Electromagnetic relays were already studied in the 1960s \cite{barkan1967study}, while solenoid actuators and on/off valves started to draw attention in the late 1980s \cite{pawlak1988transient,lim1994proportional,vaughan1996modeling}. The electromagnetic dynamics of these devices has been modeled through two different approaches. The first one is the magnetic equivalent circuit (MEC) approach, which results in low-order dynamical models well suited for control or estimation, as well as for computationally inexpensive simulations \cite{huang2022modeling,ito2019flux,ramirez2016new,zhao2020duality}. Although these models only capture the dynamics of a small number of variables, they can be made very accurate by including phenomena such as eddy currents or magnetic saturation and hysteresis \cite{ramirez2019hybrid,xu2020two}. The second approach is the use of high-order numerical models, e.g., those based on the finite element method \cite{guofu2012permanent,xu2016multiphysics}, which provide much more detailed results, but are generally computationally expensive and not appropriate for some classes of analysis. Some works combining the two methods can also be found~\cite{guofo2011output}, and mention should also be made of semi-analytical methods~\cite{gysen2010general}. Different methods have been also used to predict the motion, but the most widespread approach is the use of mass-spring-damper rigid body models with rectilinear motion and a single degree of freedom~\cite{huang2022modeling}.

Many works have been also focused on the design of control strategies to achieve soft landing, i.e., switching the device without impacts or bounces. Since the early works in the 1990s~\cite{lim1994proportional,davies1996towards}, a variety of methods have been evaluated in pursuit of this goal: sliding-mode control \cite{gamble1996comparison,EYABI2006159}, optimal open-loop control \cite{Fabbrini2012cost}, backstepping~\cite{deschaux2019magnetic}, or iterative techniques \cite{moya2020run}, among others. Many of the proposed strategies, in particular those including some form of feedback, rely on position estimators that are fed only with measurements of electrical variables \cite{STRAUBERGER2016206,braun2018observer}.
 
Although there is much research on modeling, control, and estimation of solenoid actuators, on/off valves, and electromagnetic relays, a significant gap exists in the analysis area. In this paper, we present for the first time in the literature a stability analysis for electromechanical switching devices based on a single-coil reluctance actuator. The main contribution of the work is to explain why these systems exhibit hysteretic bistable behavior with respect to the supply voltage, i.e., why they are only stable at two given positions and why the switching in one direction occurs at a different voltage level than in the other direction. The study leverages a hybrid dynamical model, based on the insights and methodologies established in our previous research~{\cite{ramirez2016new}},~{\cite{ramirez2019hybrid}}, to obtain a series of analytic expressions relating the physical parameters of the system to the switching conditions. In order to improve the understanding of the theoretical results, the expressions obtained are presented alongside plots generated using parameter values that can be considered representative of a typical electromechanical switching device. The results are discussed in depth, emphasizing the implications in the design of new switching devices and other possible practical uses. To validate the theoretical findings, we also present experimental results obtained from a real device, confirming the predicted hysteretic bistable behavior. Finally, a parameter estimation procedure based on the derived expressions demonstrates the practical value of the research.

\section{Modeling}
\label{sec:modeling}

The dynamics of the actuator is explained with the help of the diagram in Fig.~{\ref{fig:MEC}}. A coil of $N$ turns is wrapped around the core, with an electric current $i$ flowing through it. This current, whose dynamics is governed by the electrical power supply circuit, is responsible for inducing the magnetic flux $\phi$ in the core. Due to the high permeability of the iron, the magnetic flux remains confined almost entirely in the core, with the exception of the segment where it crosses the air gap. As it will be seen later, the magnetic flux is directly related to the magnetic force, which always acts by attracting both ends of the air gap. The length of the air gap, denoted by $z$ in the figure, will be used as the variable that defines the position of the actuator. Regardless of the actuator design, the air gap length is always greater than or equal to zero. In addition to this constraint---negative air gaps have no physical meaning---all actuators include some type of mechanical constraints that limit the armature stroke. Without loss of generality, it can thus be stated that $z \in \left[z_\mathrm{1},\, z_\mathrm{2}\right]$, with \mbox{$0 \leq z_\mathrm{1} < z_\mathrm{2}$}.

\subsection{Free motion dynamics}
\label{subsec:free_motion}

When the armature is strictly between the two mechanical limits, that is, $z_\mathrm{1} < z <z_\mathrm{2}$, the actuator dynamics is continuous and can be completely described by a set of ordinary differential equations. The dynamics of the entire system is the union of the mechanical dynamics, which describes the motion of the armature, and the electromagnetic dynamics, which describes how the electric current and magnetic flux evolve. The two dynamics are interconnected by means of the magnetic force.

The electromagnetic dynamics is studied first. Assuming that the coil is directly powered by a voltage source, its dynamics is governed by
\begin{equation}
u = R\, i + N \dot{\phi},
\label{eq:electrical}
\end{equation}
where $u$ is the supply voltage, $R$ is the internal resistance of the winding, and $i$ and $\phi$ are the electric current and magnetic flux as previously defined. In the design of a control system, the voltage $u$ would be considered as the controllable input. However, this paper focuses on analyzing the system under its usual mode of operation, i.e., when it is supplied with a constant voltage. Therefore, for the purposes of the subsequent analysis, $u$ is considered to be a parameter.

To find a solution to {\eqref{eq:electrical}}, an additional relationship between current and magnetic flux is required. This relationship can vary depending on the complexity of the model and the electromagnetic phenomena considered. For simplicity, this paper assumes a one-to-one relationship between current and flux, implying that magnetic hysteresis and induced currents within the core are neglected. While this simplification can lead to some discrepancies with respect to the real behavior, it allows for obtaining analytical expressions for the behavior of the system, useful for fundamental understanding, quick analyses, or design optimization. A detailed discussion of the implications of this assumption on model accuracy can be found in~{\cite{ramirez2019thesis}}.

One way to reflect this one-to-one relation is by using the reluctance of the magnetic circuit, which is a variable that depends on the geometry, the magnetic properties of the materials and, possibly, the level of magnetic excitation. Under this approach, the expression relating flux and current, known as Hopkinson's law, is
\begin{equation}
 \phi\, \mathcal{R}(z,\phi) = N\,i,
\end{equation}
where $\mathcal{R}(z,\phi)$ is the reluctance, which is in general a function of $z$ and $\phi$. A wide variety of methodologies can be found in the literature to obtain such function. Despite the detailed descriptions that can be obtained by numerical methods \cite{bao2014optimization}, the nature of the study to be performed makes analytic expressions advantageous in this case.

Two different expressions for the reluctance are used in this paper, which in turn result in two different dynamical models for the actuator. The first one is a basic model, which reflects only the fundamental dynamics of the system. In this model, the expression of the reluctance is given by
\begin{equation}
\mathcal{R}(z) = \mathcal{R}_{c0} + \mathcal{R}_{g0} + k_\mathcal{R}\,z,
\label{eq:rel_basic}
\end{equation}
where $\mathcal{R}_{c0}$, $\mathcal{R}_{g0}$ and $k_\mathcal{R}$ are strictly positive parameters. The first term, $\mathcal{R}_{c0}$, corresponds to the reluctance of the iron core, which is assumed constant. On the other hand, the other two terms model the reluctance of the air gap, which is assumed linear with respect to the gap length. Thus, $\mathcal{R}_{g0}$ is the gap reluctance for $z=0$ and $k_\mathcal{R}$ is the slope of the model. The reluctance of this model does not depend on the flux, which implies that it does not consider either magnetic hysteresis or saturation. In addition, it states that the air gap reluctance is proportional to the gap length, which means that flux fringing is ignored. Despite that, its main advantage is that it provides simple analytic expressions that relate the physical parameters to certain points of interest in the system. These expressions can be especially useful in the design of new actuators or in parameter identification procedures. 

The second model includes the phenomenon of magnetic saturation, which is known to play a significant role in the behavior of the system. This is achieved by using the Fr\"ohlich--Kennelly relation \cite{ramirez2016new}, which results in a reluctance given by
\begin{equation}
  \mathcal{R}(z,\phi) = \frac{{\mathcal{R}}_{c0}}{1-\left\vert\phi\right\vert/\phi_\mathrm{sat}}+ \mathcal{R}_{g0} + k_\mathcal{R}\,z,
\label{eq:rel_sat}
\end{equation}
where $\phi_\mathrm{sat}$ is also a strictly positive parameter representing the saturation flux and $\phi \in (-\phi_\mathrm{sat},\phi_\mathrm{sat})$. In this case, $\mathcal{R}_{c0}$ represents the reluctance of the iron core for $\phi=0$. Magnetic saturation results in an increase in the first term of the reluctance, which is the one corresponding to the core. In this sense, note that the denominator of this term varies with the flux, but is always between 0 and 1. As a result, the reluctance of this model for any position and flux is always greater than that given by \eqref{eq:rel_basic}. Note also that for $\phi_\mathrm{sat}=\infty$, i.e., when there is no saturation, this model turns into the basic one. Due to its increased sophistication, this model results in a more accurate analysis of a given actuator with known parameters \cite{ramirez2019thesis}. However, it results in expressions that are considerably more complex. Thus, its practical use in design or parameter estimation procedures is limited.

One of the benefits of using a reluctance-based model is that this variable is also directly related to the magnetic force. Given a reluctance $\mathcal{R}(z,\phi)$, this force is given \cite{ramirez2016new} by
\begin{equation}
F_\mathcal{R}(z,\phi) = -\frac{1}{2}\,\dfrac{\partial \mathcal{R}_\mathrm{}}{\partial z}\,\phi^2.
\label{eq:F_mag}
\end{equation}
Note that this expression states that the magnetic force always acts in the direction of reducing the reluctance independently of the sign of the flux. When particularized for the two reluctance models presented above, it takes the form
\begin{equation}
F_\mathcal{R}(\phi) = -\frac{1}{2}\,k_\mathcal{R}\,\phi^2.
\label{eq:F_mag2}
\end{equation}

The magnetic force is the only external excitation of the mass-spring-damper system that forms the actuator mechanism. In this work, we assume a linear spring and a damping force described by a function of the form $f_\mathrm{d}: \dot z \mapsto f_\mathrm{d}(\dot z)$, with $f_\mathrm{d}(0) = 0$. This latter assumption encompasses a wide range of frictional damping models with varying degrees of sophistication {\cite{wenzl2018comparison}}, which implies that the analyses in the following sections are more generalizable. Notably, any model that meets this assumption exhibits identical behavior at equilibrium. The dynamics of the motion, given by Newton's second law, is therefore as follows,
\begin{equation}
m \, \ddot{z} = -\frac{1}{2}\,k_\mathcal{R}\,\phi^2 - k_\mathrm{s} \left(z-z_\mathrm{s}\right) - f_\mathrm{d}(\dot{z}),
\label{eq:newton2nd}
\end{equation}
where $m$ is the armature mass and $k_\mathrm{s}$ and $z_\mathrm{s}$ are respectively the stiffness and equilibrium position of the spring. By combining all the above equations, it is possible to obtain a dynamical model of the system in state space representation. If the state vector is chosen as \mbox{$x=\left[\,z\ v\ \phi\, \right]^\intercal$}, where ${{v}}$ is the velocity, the explicit dynamics of the state variables are
\begin{align}
    &\dot{z} =f_z({x})  = v,   \label{eq:f1_stability}\\
    &\dot{v} =f_v({x}) = -\frac{k_\mathcal{R}}{2\,m}\,\phi^2  - \frac{k_\mathrm{s}}{m} \left(z-z_\mathrm{s}\right) - \frac{1}{m}\, f_\mathrm{d}(\dot{z}),\label{eq:f2_stability}\\
    &\dot{\phi} =f_\phi({x},u)  = \frac{u}{N}-\frac{R}{N^2}\, \phi\,\mathcal{R}(z,\phi).\label{eq:f3_stability}
\end{align}
The dependence of $f_\phi$ on $u$ has been decided to be explicitly stated because of the importance of the latter, regardless of whether it is considered the input or a mere parameter.

\subsection{Dynamics of switching actuators}
\label{subsec:dyn_switch}

Although there are some reluctance actuators designed to operate in continuous mode \cite{vrijsen2014prediction}, all low-cost switching devices based on this technology are intended to switch between the two limit positions. This category includes electromagnetic relays, whose purpose is to open and close electrical connections, and solenoid actuators, which are widely used as on/off valves in hydraulic and pneumatic circuits. 
The mechanical limits of these devices not only constrain the movement of the armature to a certain range, but also result in the occurrence of strong impacts at the end of the displacements. Furthermore, these impacts often lead to armature bouncing. Since the velocity changes instantaneously on impacts, this behavior can be considered as a case of hybrid dynamics.

\begin{figure}[t]
\centering
\includegraphics[width=7.5cm]{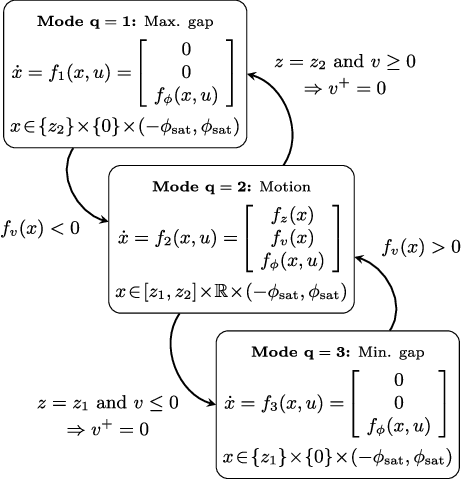}%
\caption{Hybrid automaton modeling the hybrid dynamics of a switching actuator. If the model equations do not include saturation, $\phi_\mathrm{sat}=\infty$.}
\label{fig:hybrid_automaton}
\end{figure}

It is thus clear that the differential equations obtained in the previous section are not sufficient to describe the dynamics of these switching devices. For this purpose, the hybrid automaton shown in Fig.~\ref{fig:hybrid_automaton} will be used. This automaton considers three different dynamic modes denoted by variable $q \in Q=\{1,2,3\}$. One of them ($q=2$) models the dynamics during the armature motion and the others ($q=1,\,3$) describe the dynamic behavior of the system when the armature is at rest at the extreme positions. The functions $f_z$, $f_v$, and $f_\phi$ correspond to those in \eqref{eq:f1_stability}--\eqref{eq:f3_stability}. The domain of the state in each dynamic mode is also indicated below the dynamics. Transitions between modes are indicated by arrows next to which is shown the condition that must be met to produce the jump. If the transition involves an instantaneous change in the value of one of the state variables, this is indicated after a rightwards double arrow ($\Rightarrow$). The superscript $^+$ is used to denote the value after the jump. 

Following the procedure described by \cite{goebel2009hybrid}, this hybrid automaton can be easily cast into the more general form
\begin{align}
    \dot{x} =&\ f_{q}(x,u), & q \in Q,\ &x \in C_q, \label{eq:hybrid_autom_3}\\
    (x^+,q^+) =&\ g_{q}(x),  & q \in Q,\ &x \in D_q, \label{eq:hybrid_autom_4}
\end{align}
where $C_q$, $D_q$, and $g_q$, for all $q \in Q$, can be directly obtained from Fig.~\ref{fig:hybrid_automaton} as
\begin{align}
    &C_1 = \{z_\mathrm{2}\} \!\times\! \{ 0 \} \!\times\! (-\phi_\mathrm{sat},\phi_\mathrm{sat}),\nonumber \\
    &C_2 = [z_\mathrm{1},z_\mathrm{2}] \!\times\! \mathbb{R} \!\times\! (-\phi_\mathrm{sat},\phi_\mathrm{sat}),\nonumber \\
    &C_3 = \{z_\mathrm{1}\} \!\times\! \{ 0 \} \!\times\! (-\phi_\mathrm{sat},\phi_\mathrm{sat}),\nonumber \\
    &D_1 = \{x \mid f_v({x})<0\},\nonumber \\
    &D_2 = D_{2,1} \cup D_{2,3},\nonumber \\
    &D_{2,1} = \{z_\mathrm{2}\} \!\times\! \mathbb{R}_{\geq 0} \!\times\! (-\phi_\mathrm{sat},\phi_\mathrm{sat}),\nonumber \\
    &D_{2,3} = \{z_\mathrm{1}\} \!\times\! \mathbb{R}_{\leq 0} \!\times\! (-\phi_\mathrm{sat},\phi_\mathrm{sat}),\nonumber \\
    &D_3 =  \{x \mid f_v({x})>0\},\nonumber \\
    &g_1(x)= (x,2),\nonumber \\
    &g_2(x) = \begin{cases} 
      (\left[\,z\ \, 0\ \, \phi\, \right]^\intercal,1) & x \in D_{2,1}, \\
      (\left[\,z\ \, 0\ \, \phi\, \right]^\intercal,3) & x \in D_{2,3}, 
   \end{cases} \nonumber \\
    &g_3(x) = (x,2). \nonumber
\end{align}
This formulation will be used later when studying the stability in the hybrid case.

As a final remark, it should be noted that the impacts modeled by this automaton are purely inelastic and, therefore, this model is not able to reproduce the bouncing phenomenon. While incorporating such phenomenon would enrich the model, it is important to note that its transient effects do not affect the location of the equilibrium points. Therefore, it is not necessary to consider such bounces in order to study the stability of the system. In any case, the model could easily be modified to include this phenomenon if desired, i.e., by using a restitution coefficient for the velocity~{\cite{da2022compendium}}.

\section{Analysis in continuous mode}\label{sec:analysis_cont}

In this and the following section, we analyze the equilibrium points and the stability of the system as functions of the supply voltage. As it will be seen, the variation of $u$ produces certain changes in the phase space that explain to a large extent the behavior of this class of devices. We present, whenever possible, the closed-form expressions of the points of interest as a function of the parameters. The figures, included to improve the understanding of the theoretical results, have been generated using the parameters in Table~{\ref{tab:param_values}}, which can be considered representative of many commercial solenoid~actuators.

As discussed in the previous section, electromechanical switching devices exhibit hybrid dynamics due to mechanical constraints that limit the armature stroke. In this section, however, a preliminary analysis is presented in which the system is assumed to be always in the dynamic mode $q=2$ (see Fig.~{\ref{fig:hybrid_automaton}}). The goal is to provide a basis for the analysis presented in the next section, where all operating modes of the actuator are considered. However, this analysis may be useful on its own for certain high-precision actuators, such as those used in the semiconductor industry~{\cite{vrijsen2014prediction}}, that are specifically designed to operate only in continuous mode. In this section, the dynamics is assumed to be continuous and given by
\begin{equation}
    \dot{x} = f_2(x,u) = \left[\begin{array}{ccc}f_z(x) & f_v(x) & f_\phi(x,u) \end{array}\right]^\intercal,
    \label{eq:cont_prop_fx}
\end{equation}
where $x \in [z_\mathrm{1},z_\mathrm{2}] \times \mathbb{R} \times (-\phi_\mathrm{sat},\phi_\mathrm{sat})$. It is also assumed that there is no limit for the maximum position of the armature. That is, $z_\mathrm{2} = \infty$. However, since negative air gaps do not make physical sense, we do restrict the analysis to the limit case $z_\mathrm{1}=0$. The considered equilibria are therefore those points that, being inside the domain, satisfy \mbox{$f_2(x,u)=0$}. In particular, the condition $f_z(x)=v=0$ implies that the velocity at all these points must be equal to zero. For this reason, in the following analyses it is implicitly assumed that $v=0$ and, hence, only the $z$- and $\phi$-coordinates of the equilibria are analyzed and discussed.

\begin{table}[t]
	\renewcommand{\arraystretch}{1.3}
	\caption{Parameter values used for graphical representation.\\The parameter $\phi_\mathrm{sat}$ is only used when considering the model with magnetic saturation.}
	\centering
    \label{tab:param_values}
	\begin{tabular}{ccccc}
		\cmidrule{1-2} \cmidrule{4-5}
		Parameter & Value &  &
		Parameter & Value \\ 
		\cmidrule{1-2} \cmidrule{4-5}
		$R$ & { $50$} $\Omega$ &&
		$N$ & { $1200$}       \\
        $\mathcal{R}_{c0}$ & { $1.5\!\cdot\!10^{7}$ H$^{-1}$} &&
        $m$ & {$1$ g}\\
        $\mathcal{R}_{g0}$ & { $0$ H$^{-1}$} &&
        $k_\mathrm{s}$ & {$55$ N/m}\\
        $k_\mathcal{R}$ & { $2\!\cdot\!10^{10}$ H$^{-1}$/m} &&
        $z_\mathrm{s}$ & {$15$ mm}\\
        $\phi_\mathrm{sat}$ & { $20$ $\upmu$Wb} &&
        - & -\\
		\cmidrule{1-2} \cmidrule{4-5} 
	\end{tabular}
\end{table}

\subsection{Basic model}

\begin{figure}[t]
\centering
    \includegraphics[]{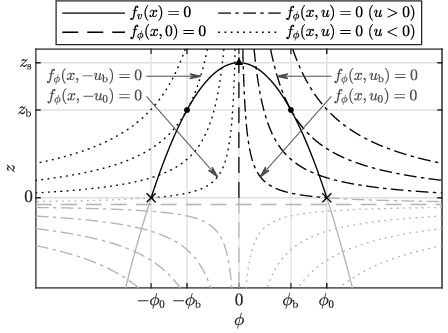}%
    \caption{Basic model. Curves $f_v(x)=0$ and $f_\phi(x,u)=0$ in the $z$-$\phi$ plane for different values of $u$. As the absolute value of $u$ increases, the hyperbolas $f_\phi(x,u)=0$ move farther away from $f_\phi(x,0)=0$. Values outside the domain of the functions are shown in light gray.}
    \label{fig:equilibrium_state_2}
\end{figure}

The basic model is studied first. Recall that this model considers the reluctance given by \eqref{eq:rel_basic}, which leads to the magnetic force \eqref{eq:F_mag2}. Geometrically, the equilibrium points of the system on the $z$-$\phi$ plane are points of intersection of the curves $f_v(x)=0$ and $f_\phi(x,u)=0$.  Fig.~\ref{fig:equilibrium_state_2} depicts these curves for different values of $u$. It is shown that, depending on the supply voltage, one, two, or three intersection points may exist. Note however that some of them are outside the domain and, thus, are not equilibrium points of the system.

Let us begin the analysis with the particular case $u=0$. For this supply voltage there are three points at which the parabola $f_v(x)=0$ intersects the two straight lines resulting from $f_\phi(x,0)=0$, i.e, $\phi=0$ and $z=-\mathcal{R}_0/k_\mathcal{R}$. Only the point at $(\phi,z)=(0,z_\mathrm{s})$---indicated with a black triangle in the figure---is inside the domain and, therefore, there is only one equilibrium point in the system. Note that this point coincides with the equilibrium position of the spring. 

Then, as the absolute value of $u$ increases, the two straight lines become a hyperbola, which initially also intersects the parabola at three points. For small supply voltages, the picture is similar to the case $u=0$, i.e., only one of the points corresponds to a positive air gap ($z\geq 0$). However, there is a threshold absolute value of $u$---denoted as $u_0$---that brings another intersection point into the domain. This value can be obtained from the solution to the system of equations consisting of $f_2(x,u)=0$ and $z=0$. There are two solutions, one for positive supply voltage $u=u_0$ at $(\phi,z)=(\phi_0,0)$, and the other one for negative voltage $u=-u_0$ at $(\phi,z)=(-\phi_0,0)$. Both are indicated with crosses in the figure. The values of $u_0$ and $\phi_0$, which depend on the model parameters, are
\begin{align}
        &u_{0} = \dfrac{R\,\phi_{0}}{N}\,\left(\mathcal{R}_{c0} + \mathcal{R}_{g0} \right), \label{eq:u0}\\
    &\phi_{0} = \sqrt{\dfrac{2\, k_\mathrm{s}\, z_\mathrm{s}}{k_\mathcal{R}}}. \label{eq:phi0}
\end{align}

Then, for supply voltages greater in absolute value than $u_0$, there are two other critical points---marked as dots in the figure---at which the parabola and the hyperbola become tangent. These are tangential bifurcation points where the two existing equilibrium points converge. They can be calculated analytically, either geometrically or by taking into account that the Jacobian vanishes at these points. That is, they can be obtained from the solution of the system of equations formed by $f(x,u)=0$ and $\vert{\partial f}/{\partial x}\vert=0$. There are two solutions as in the previous case: one for positive supply voltage $u=u_\mathrm{b}$ at $(\phi,z)=(\phi_\mathrm{b},z_\mathrm{b})$, and another one for negative voltage $u=-u_\mathrm{b}$ at $(\phi,z)=(-\phi_\mathrm{b},z_\mathrm{b})$. The expressions of $u_\mathrm{b}$, $z_\mathrm{b}$, and $\phi_\mathrm{b}$ are as follows:
\begin{align}
        &u_\mathrm{b} = \dfrac{2\,R\,\sqrt{6\,k_\mathrm{s}\left(\mathcal{R}_{c0} + \mathcal{R}_{g0} +k_\mathcal{R}\,z_\mathrm{s}\right)^3}}{9\,N\,k_\mathcal{R}},\\
    &z_\mathrm{b} = \dfrac{2}{3}\, z_\mathrm{s} - \dfrac{\mathcal{R}_{c0} + \mathcal{R}_{g0} }{3\,k_\mathcal{R}}, \label{eq:zb}\\ 
    &\phi_\mathrm{b} = \dfrac{\displaystyle\sqrt{6\,k_\mathrm{s}\left(\mathcal{R}_{c0} + \mathcal{R}_{g0} +k_\mathcal{R}\,z_\mathrm{s}\right)}}{3\,k_\mathcal{R}}.
\end{align}
Finally, for $\lvert u \rvert>u_\mathrm{b}$ the curves $f_v(x)=0$ and $f_\phi(x,u)=0$ intersect only at one point, which is always outside the domain. Thus, there are no equilibrium points for supply voltages beyond that value.

\begin{figure}[t]
\centering
    \includegraphics[]{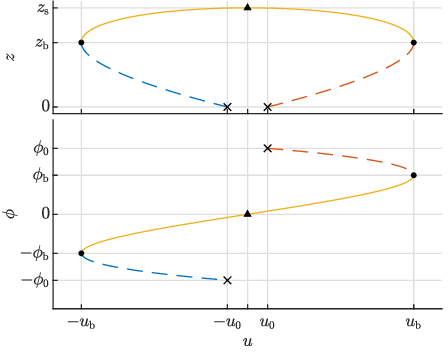}%
    \caption{Basic model under continuous operation. Bifurcation diagram of equilibria in terms of $u$. Position (top) and magnetic flux (bottom). Stable/unstable equilibria are plotted with \mbox{solid/dashed} lines.}
    \label{fig:equilibrium_state_1}
\end{figure}

Additional results are represented in Fig.~\ref{fig:equilibrium_state_1}. This figure depicts the $z$- and $\phi$-coordinates of the equilibrium points as functions of $u$. Stability is indicated in the usual way: stable points are represented by solid lines and unstable points by dashed lines. Stability has been analyzed through Lyapunov's indirect method, i.e., by checking whether all eigenvalues of the Jacobian matrix ${\partial f}/{\partial x}$ lie in the left half complex plane. The figure shows the traces of the three potential equilibrium points previously analyzed. Due to the domain of the functions, two of them (red and blue lines) do not coexist for any value of $u$. These equilibria are unstable and symmetric to each other with respect to $(\phi,u)=(0,0)$. The third equilibrium (yellow line) is stable, but only exists for a limited range of positions, $z \in \left[z_\mathrm{b}, z_\mathrm{s}\right]$. As already stated, it corresponds exactly to the equilibrium position of the spring when $u=0$. All the points of interest previously discussed are also indicated in the figure: the spring equilibrium position is denoted with a triangle, the saddle-node bifurcation points are marked as dots and the points where the unstable equilibria escape from the domain are marked with crosses.

\subsection{Model with magnetic saturation}

The model with saturation has the same dynamics as the basic model except for the reluctance expression, which takes the form \eqref{eq:rel_sat}. This reluctance results exactly in the same magnetic force, so $f_v$ remains unchanged. Thus, only the function $f_\phi$ is altered. The analysis presented below follows the same line of reasoning as in the previous case. That is, we study the intersection points of the curves $f_v(x)=0$ and $f_\phi(x,u)=0$ and discuss under what conditions they are equilibrium points of the system. In the analysis it is assumed that $\phi_\mathrm{sat}>\phi_0$, where $\phi_0$, which is given by \eqref{eq:phi0}, is the flux value at which the curve $f_v(x)=0$ intersects the $z=0$ axis. 

\begin{figure}[t!]
\centering
    \includegraphics[]{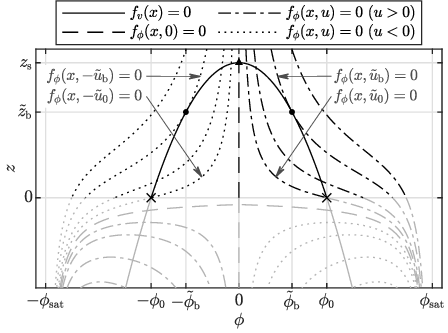}%
    \caption{Model with magnetic saturation. Curves $f_v(x)=0$ and $f_\phi(x,u)=0$ in the $z$-$\phi$ plane for different values of $u$. As the absolute value of $u$ increases, the curves $f_\phi(x,u)=0$ move farther away from $f_\phi(x,0)=0$. Values outside the domain of the functions are shown in light gray.}
    \label{fig:equilibrium_state_2_sat}
\end{figure}

As seen in Fig.~\ref{fig:equilibrium_state_2_sat}, magnetic saturation significantly modifies the curves {$f_\phi(x,u)=0$}. The most notable difference is the presence of two asymptotes at the flux saturation values, which cause these curves and the parabola {$f_v(x)=0$} to actually intersect at more than three points. These new intersections---which are not represented in the figure---are nevertheless outside the domain, so they do not give rise to new equilibria. The equilibrium analysis is, in fact, qualitatively similar to that of the basic model. For $u=0$ there is only one equilibrium point, which is the equilibrium position of the spring. By increasing the absolute value of $u$, one of the intersection points enters the domain and gives rise to a second equilibrium point. Since the parabola $f_v(x)=0$ has not changed, this point is at the same location as in the basic model, but in this case it occurs for a different value of the supply voltage. Again, there are two symmetric solutions, one for positive supply voltage $u=\tilde{u}_0$ at $(\phi,z)=(\phi_0,0)$, and the other one for negative voltage $u=-\tilde{u}_0$ at $(\phi,z)=(-\phi_0,0)$. Note that the specific results of the model with saturation will be denoted with tildes. The supply voltage $\tilde{u}_0$ for which these equilibria occur is given by
\begin{equation}
        \tilde{u}_{0}= \dfrac{R\,\phi_{0}}{N}\,\left(\frac{\mathcal{R}_{c0}}{1-{\phi_{0}}/{\phi_\mathrm{sat}}}+\mathcal{R}_{g0} \right). \label{eq:u0tilde}
\end{equation} 
Considering that $\mathcal{R}_{c0}>0$, $\mathcal{R}_{g0}>0$ and $\phi_\mathrm{sat}>\phi_0>0$, it is straightforward that $\tilde{u}_{0}$ is greater than the voltage $u_0$ given by {\eqref{eq:u0}}. In other words, the supply voltage required to reach the aforementioned equilibrium point is higher when magnetic saturation is taken into account.%

In fact, considering that the reluctance of the model with saturation is always higher than that of the basic model, it can be shown that the voltage required to reach any given equilibrium point is always higher when considering saturation.

The tangential bifurcations that exist in the basic model are also present in this case. These occur for those values of $u$ at which the curves $f_\phi(x,u)=0$ and $f_v(x)=0$ become tangent. There are two solutions: one for positive supply voltage $u=\tilde{u}_\mathrm{b}$ at $(\phi,z)=(\tilde{\phi}_\mathrm{b},\tilde{z}_\mathrm{b})$, and another one for negative voltage $u=-\tilde{u}_\mathrm{b}$ at $(\phi,z)=(-\tilde{\phi}_\mathrm{b},\tilde{z}_\mathrm{b})$. In this case, it is also possible to obtain closed-form solutions for these points, but the resulting expressions are of such length that their practical use is limited. In fact, we do not present them in the manuscript for the sake of brevity. Comparing the curves in Figs.~\ref{fig:equilibrium_state_2} and~\ref{fig:equilibrium_state_2_sat}, however, it can be seen that the bifurcations in this model must necessarily occur at points farther away from the vertex of the parabola. That is, $\tilde{z}_\mathrm{b} < {z}_\mathrm{b}$ and $\tilde{\phi}_\mathrm{b}>{\phi}_\mathrm{b}$. The relation between $\tilde{u}_\mathrm{b}$ and ${u}_\mathrm{b}$, which can be obtained by solving $f_\phi(x,u)=0$ for each of the models, is given by
\begin{equation}
        \dfrac{\tilde{u}_\mathrm{b}}{{u}_\mathrm{b}}= \dfrac{\tilde{\phi}_\mathrm{b}}{{\phi}_\mathrm{b}}\,\dfrac{\dfrac{{\mathcal{R}}_{c0}}{1-\tilde{\phi}_\mathrm{b}/\phi_\mathrm{sat}} + \mathcal{R}_{g0} + k_\mathcal{R}\,\tilde{z}_\mathrm{b}}{\mathcal{R}_{c0} + \mathcal{R}_{g0} + k_\mathcal{R}\,{z}_\mathrm{b}},
\end{equation}
and it is not possible to determine a priori which of the two is greater than the other.

\begin{figure}[t]
\centering
    \includegraphics[]{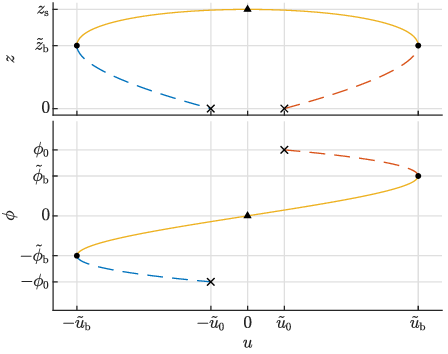}%
    \caption{Model with saturation under continuous operation. Bifurcation diagram of equilibria in terms of $u$. Position (top) and magnetic flux (bottom). Stable/unstable equilibria are plotted with \mbox{solid/dashed} lines.}
    \label{fig:equilibrium_state_1_sat}
\end{figure}

Fig.~\ref{fig:equilibrium_state_1_sat} shows the $z$- and $\phi$-coordinates of the equilibria as functions of $u$. It is shown that, despite the above-mentioned differences, the relationship between the equilibrium points and the supply voltage in this model is qualitatively similar to that obtained for the basic model (see Fig.~\ref{fig:equilibrium_state_1}). Equivalently to the previous case, note that there are no equilibrium points for supply voltages greater than $\tilde{u}_\mathrm{b}$ in absolute value.

\subsection{Discussion}

The analysis presented in this section is based on the assumption that the device is always operating in the dynamic mode $q=2$. Although this situation does not usually occur in real actuators, some interesting conclusions can be drawn from the above results with respect to a potential position controller. It has been shown that it is impossible to reach equilibrium at a position higher than the spring equilibrium position. This, which is a consequence of the fact that the magnetic force is always attractive, is an important consideration to take into account in the actuator design. In addition, it has also been shown that open-loop stabilization of this class of actuators is only possible at positions close to the spring equilibrium position. Thus, if the position is to be stabilized for some $z<z_\mathrm{b}$---or $z<\tilde{z}_\mathrm{b}$, when considering saturation---it is imperative to implement some form of feedback. Note that the value of $\tilde{z}_\mathrm{b}$ is upper bounded by ${z}_\mathrm{b}$, which is related to the physical parameters according to \eqref{eq:zb}. This information can be used during the actuator design process to estimate the region in which the actuator will be open-loop stable.

\section{Analysis under hybrid dynamics}

In this section, we extend the previous stability analysis by considering that the device may actually operate in any of the dynamic modes of the hybrid automaton in Fig.~\ref{fig:hybrid_automaton}, i.e., moving freely between the ends of the stroke or at rest in one of the two travel limits. As a result, we consider that the motion of the armature is bounded between $z_\mathrm{1}\geq 0$ and $z_\mathrm{2}>z_\mathrm{1}$. The hybrid dynamics is considered hereafter in the analysis to study the effects of these position boundaries. Note that the equilibrium points of a hybrid system are those locations at which the state neither evolves continuously nor discretely. Hence, a point $(x,q)$ is an equilibrium  of the hybrid automaton \eqref{eq:hybrid_autom_3}--\eqref{eq:hybrid_autom_4} if
\begin{equation}
   f_{q}(x,u)=0, \hspace{1em} q \in Q,\ x \in C_{q}\cap \overline{D_{q}}.
   \label{eq:eq_points_hybrid}
\end{equation}
Since this automaton has three dynamic modes, the equilibrium points of the system are therefore the solutions of \eqref{eq:eq_points_hybrid} for $q\in Q=\left\{1,2,3\right\}$.

It is not difficult to realize that this analysis is much richer than in the continuous case, as the behavior of the system may change considerably depending on the values of $z_\mathrm{1}$ and $z_\mathrm{2}$. In this work, we assume that $z_\mathrm{1} \in \left[ 0, {z}_\mathrm{b}\right)$---or $z_\mathrm{1} \in \left[ 0, \tilde{z}_\mathrm{b}\right)$, when considering saturation---a condition that is met in virtually all commercial devices. In the following, we analyze, for the two models previously proposed, the different cases that may arise depending on the value of $z_\mathrm{2}$.

\subsection{Basic model}

\begin{figure}[t]
\centering
    \subfloat[]{
    \includegraphics[]{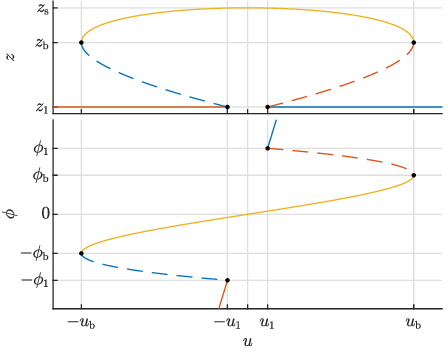}
    \label{equilibrium_state_hybrid_1}}\\
    \subfloat[]{
    \includegraphics[]{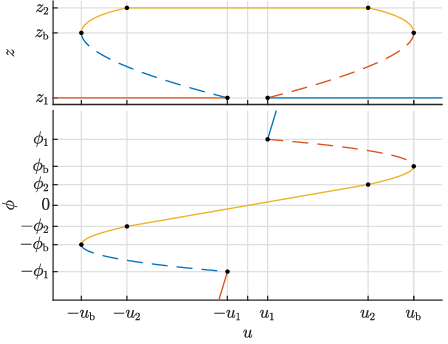}
    \label{equilibrium_state_hybrid_2}}\\
    \subfloat[]{
    \includegraphics[]{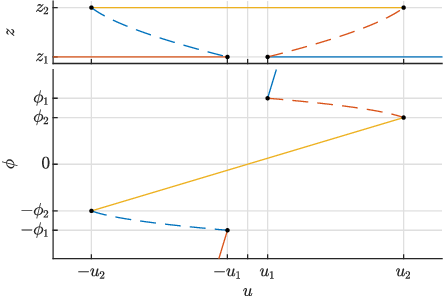}
    \label{equilibrium_state_hybrid_3}}%
    \caption{Basic model under hybrid dynamics. Bifurcation diagram of equilibria in terms of $u$. Position (top) and magnetic flux (bottom). Stable/unstable equilibria are plotted with \mbox{solid/dashed} lines. (a)~Case~1: $z_\mathrm{2}>z_\mathrm{s}$. (b)~Case~2: $z_\mathrm{b}<z_\mathrm{2}<z_\mathrm{s}$. (c)~Case~3: $z_\mathrm{1}<z_\mathrm{2}<z_\mathrm{b}$.}
    \label{fig:equilibrium_state_hybrid}
\end{figure}

The results corresponding to the basic model are presented in Fig.~\ref{fig:equilibrium_state_hybrid}. There are three cases depending on the value of $z_\mathrm{2}$. Although not very common in practice, the first case (Fig.~\ref{equilibrium_state_hybrid_1}) corresponds to a maximum position greater than the equilibrium position of the spring, i.e., $z_\mathrm{2}>z_\mathrm{s}$. It can be seen that there is only one equilibrium for $u=0$, which corresponds to the spring equilibrium position, and also that this equilibrium has exactly the same behavior as in the continuous case. As $u$ increases, it approaches $(z,\phi)=(z_\mathrm{b},\phi_\mathrm{b})$ and, for $u>u_\mathrm{b}$, it vanishes. Equivalent results occur for negative supply voltages. More interesting is the behavior of the other two points. It is shown that there exists a stable point at the minimum position ($z=z_\mathrm{1}$) for supply voltages $u \in \left(-\infty,-u_\mathrm{1}\right]\cup\left[u_\mathrm{1},+\infty\right)$, where $u_\mathrm{1}$ is obtained from the intersection of $f_v(x)=0$ and $f_\phi(x,u)=0$ when $z=z_\mathrm{1}$.
\begin{equation}
    u_\mathrm{1} = \dfrac{R\left(\mathcal{R}_{c0} + \mathcal{R}_{g0} + k_\mathcal{R}\, z_\mathrm{1}\right)}{N}\sqrt{\dfrac{2\, k_\mathrm{s} \left(z_\mathrm{s}-z_\mathrm{1}\right)}{k_\mathcal{R}}}
    \label{eq:u1}
\end{equation}
This stable point, which does not appear if only the continuous dynamics is considered, explains why the mover and the stator of a reluctance actuator stay together at $z=z_\mathrm{1}$ if the input is high enough in absolute value. In this regard, note that the expression \eqref{eq:u1} gives the absolute value of the supply voltage at which the mover takes off from the minimum position. The corresponding magnetic flux, which is obtained directly from $f_v(x)=0$ when $z=z_\mathrm{1}$ and denoted as $\phi_\mathrm{1}$, is given by
\begin{equation}
    \phi_\mathrm{1}  =\sqrt{\dfrac{2\, k_\mathrm{s} \left(z_\mathrm{s}-z_\mathrm{1}\right)}{k_\mathcal{R}}}.
    \label{eq:phi1}
\end{equation}
Finally, note that the third equilibrium is unstable and varies its position between $z_\mathrm{1}$ and $z_\mathrm{b}$. Thus, it cannot be exploited in open-loop.

The second case (Fig.~\ref{equilibrium_state_hybrid_2}) arises when the maximum position of the mover is smaller than the spring equilibrium position and, at the same time, greater than $z_\mathrm{b}$, i.e., \mbox{$z_\mathrm{b}<z_\mathrm{2}<z_\mathrm{s}$}. As shown in the figure, this case is identical to the first one except for the first equilibrium point, which stays at $z=z_\mathrm{2}$ for supply voltages $u \in \left[-u_\mathrm{2},u_\mathrm{2} \right]$. The expression for $u_\mathrm{2}$ can be obtained as the solution to $f_v(x)=0$ and $f_\phi(x,u)=0$ when $z=z_\mathrm{2}$.
\begin{equation}
    u_\mathrm{2} = \dfrac{R\left(\mathcal{R}_{c0} + \mathcal{R}_{g0} + k_\mathcal{R} z_\mathrm{2}\right)}{N}\sqrt{\dfrac{2\, k_\mathrm{s} \left(z_\mathrm{s}-z_\mathrm{2}\right)}{k_\mathcal{R}}}
    \label{eq:u2}
\end{equation}
Analogously to \eqref{eq:u1}, expression \eqref{eq:u2} gives the absolute value of the supply voltage at which the armature lifts off from $z_\mathrm{2}$. The magnetic flux in such situation, denoted as $\phi_\mathrm{2}$, is given by
\begin{equation}
    \phi_\mathrm{2}  =\sqrt{\dfrac{2\, k_\mathrm{s} \left(z_\mathrm{s}-z_\mathrm{2}\right)}{k_\mathcal{R}}}.
    \label{eq:phi2}
\end{equation}

The third and last case (Fig.~\ref{equilibrium_state_hybrid_3}) is by far the most common in commercial switching devices and corresponds to a maximum position $z_\mathrm{2}$ such that $z_\mathrm{1} < z_\mathrm{2}<z_\mathrm{b}$. Although similar to the previous cases, the distinctive feature of this situation is that stable equilibria exist only at the limits of motion, i.e., at the discrete positions $z=z_\mathrm{1}$ and $z=z_\mathrm{2}$.

\subsection{Model with magnetic saturation}

\begin{figure}[t]
\centering
    \subfloat[]{\includegraphics[]{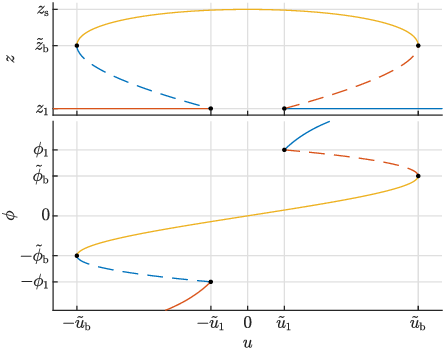}
    \label{equilibrium_state_hybrid_1_sat}}\\
    \subfloat[]{\includegraphics[]{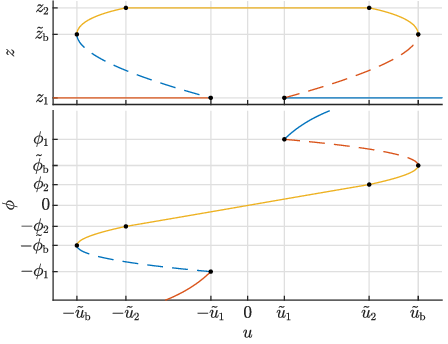}
    \label{equilibrium_state_hybrid_2_sat}}\\
    \subfloat[]{\includegraphics[]{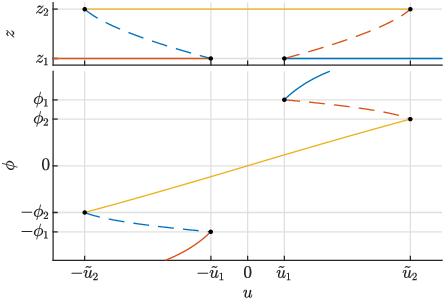}
    \label{equilibrium_state_hybrid_3_sat}}%
    \caption{Model with saturation under hybrid dynamics. Bifurcation diagram of equilibria in terms of $u$. Position (top) and magnetic flux (bottom). Stable/unstable equilibria are plotted with \mbox{solid/dashed} lines. (a)~Case~1: $z_\mathrm{2}>z_\mathrm{s}$. (b)~Case~2: $\tilde{z}_\mathrm{b}<z_\mathrm{2}<z_\mathrm{s}$. (c)~Case~3: $z_\mathrm{1}<z_\mathrm{2}<\tilde{z}_\mathrm{b}$.}
    \label{fig:equilibrium_state_hybrid_sat}
\end{figure}

In the analysis of Section~\ref{sec:analysis_cont} it has been discussed why the model with saturation, despite using a different expression for the reluctance, is qualitatively similar to the basic model in terms of equilibrium points and stability. These similarities also occur when considering the hybrid dynamics, as can be seen from the results shown in Fig.~\ref{fig:equilibrium_state_hybrid_sat}. In these plots, $\phi_\mathrm{1}$ and $\phi_\mathrm{2}$ are still given by \eqref{eq:phi1} and \eqref{eq:phi2}, respectively. This is a consequence of both models having the same expression for $f_v(x)$. On the other hand, the supply voltages at which the armature lifts off from the limit positions, which are denoted as $\tilde{u}_\mathrm{1}$ and $\tilde{u}_\mathrm{2}$, have different expressions in this case. These can be obtained as the solution to $f_v(x)=0$ and $f_\phi(x,u)=0$ at $z=z_\mathrm{1}$ or $z=z_\mathrm{2}$.
\begin{align}
    \tilde{u}_\mathrm{1} &= \dfrac{R\, \phi_\mathrm{1}}{N} \left(\frac{{\mathcal{R}}_{c0}}{1-\phi_\mathrm{1}/\phi_\mathrm{sat}} + {\mathcal{R}}_{g0} + k_\mathcal{R}\, z_\mathrm{1} \right)
    \label{eq:tilde_u1}\\
    \tilde{u}_\mathrm{2} &= \dfrac{R\, \phi_\mathrm{2}}{N} \left(\frac{{\mathcal{R}}_{c0}}{1-\phi_\mathrm{2}/\phi_\mathrm{sat}} + {\mathcal{R}}_{g0} + k_\mathcal{R}\, z_\mathrm{2} \right)
    \label{eq:tilde_u2}
\end{align}
Assuming that $\phi_\mathrm{2}<\phi_\mathrm{1}<\phi_\mathrm{sat}$---a reasonable assumption because otherwise the actuator would not be able to switch---it can be easily verified that $\tilde{u}_\mathrm{1}>{u}_\mathrm{1}$ and $\tilde{u}_\mathrm{2}>{u}_\mathrm{2}$. That is, the supply voltages at which the device switches are higher when magnetic saturation is considered.
As a summary, Table~{\ref{tab:symbols}} shows all the symbols used in the stability analysis.

\begin{table}[t]
	\renewcommand{\arraystretch}{1.3}
    \caption{List of symbols used in the stability analysis.}
    \centering
    \label{tab:symbols}
    \begin{tabular}{c@{\hspace{1em}}l}
        \hline
        Symbol &  Description\\
         \hline
        $\phi_0$ &  Magnetic flux for equilibrium at $z=0$ \\
        $u_0$ &  Voltage for equilibrium at $z=0$ (Basic model)\\
        $\tilde{u}_0$ &  Voltage for equilibrium at $z=0$ (Model with saturation)\\
        \hline
        $z_\mathrm{b}$ & Position at the bifurcation point (Basic model)\\
        $\phi_\mathrm{b}$ &  Magnetic flux at the bifurcation point (Basic model)\\
        $u_\mathrm{b}$ &  Voltage at the bifurcation point (Basic model)\\
        $\tilde z_\mathrm{b}$ & Position at the bifurcation point (Model with saturation)\\
        $\tilde{\phi}_\mathrm{b}$ &  Magnetic flux at the bifurcation point (Model with saturation)\\
        $\tilde{u}_\mathrm{b}$ &  Voltage at the bifurcation point (Model with saturation)\\
        \hline
        $z_1$ & Position corresponding to minimum air gap\\
        $\phi_1$ &  Magnetic flux for equilibrium at $z=z_1$\\
        $u_1$ &  Voltage for equilibrium at $z=z_1$ (Basic model)\\
        $\tilde{u}_1$ &  Voltage for equilibrium at $z=z_1$ (Model with saturation)\\
        \hline
        $z_2$ & Position corresponding to maximum air gap\\
        $\phi_2$ &  Magnetic flux for equilibrium at $z=z_2$\\
        $u_2$ &  Voltage for equilibrium at $z=z_2$ (Basic model)\\
        $\tilde{u}_2$ &  Voltage for equilibrium at $z=z_2$ (Model with saturation)\\
        \hline
    \end{tabular}
    \label{tab:my_label}
\end{table}

\subsection{Discussion}

\begin{figure}[t]
\centering
    \includegraphics[]{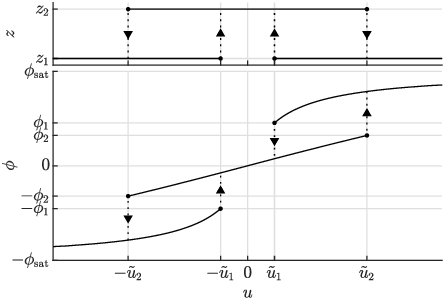}%
    \caption{Hysteretic switching loop for the model with magnetic saturation ($z_{1}<z_{2}<z_\mathrm{b}$).}
    \label{fig:hysteresis_hybrid_sat}
\end{figure}

The stability results presented in this section explain to a large extent the switching behavior of devices such as electromagnetic relays or solenoid actuators. If we focus on the third case, which is the most common in practice, it has been shown that this class of actuators are open-loop stable only at the boundary positions. In essence, this is an analytical explanation for the bistable behavior of these devices. In addition, they suffer from a distinctive hysteretic switching behavior that can also be explained from the results. Note that, when these devices are not supplied with power, the armature rests at the maximum position. If the supply voltage is steadily increased up to $u_\mathrm{2}$---or $\tilde{u}_\mathrm{2}$---the armature will suddenly switch from $z_\mathrm{2}$ to $z_\mathrm{1}$. Then, if the voltage is reduced down to $u_\mathrm{1}$---or $\tilde{u}_\mathrm{1}$---the mover will take off from $z_\mathrm{1}$ and move back up again to $z_\mathrm{2}$. This hysteresis cycle is graphically represented in Fig.~\ref{fig:hysteresis_hybrid_sat} for the model with saturation. This figure is derived directly from the results in Fig.~\ref{equilibrium_state_hybrid_3_sat}. Thus, the already presented results would permit to obtain the hysteresis loop for any other case.

It is important to note that the observed hysteretic behavior between voltage, magnetic flux, and position is independent of magnetic hysteresis, which is not included in this study for the reasons cited in Section~{\ref{subsec:free_motion}}. This implies that these devices can exhibit two forms of hysteresis simultaneously: the steady-state voltage-flux-position hysteresis studied in this work and the magnetic hysteresis between current and magnetic flux.

\section{Experimental demonstration}\label{sec:experimental}

To complement the theoretical analysis, this section presents experimental results aimed at demonstrating the applicability of the analysis to practical scenarios. The experiments address two main goals. First, to show that it is possible to replicate experimentally the theoretical hysteretic loop depicted in Fig.~{\ref{fig:hysteresis_hybrid_sat}} using measurements from a real device. Secondly, to explain how to use the measured data to estimate the key parameters that govern the behavior of such device.

\begin{figure}[t]
    \centering
    \subfloat[]{\includegraphics[width=0.39\linewidth]{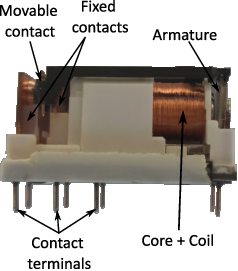}%
    \label{fig:relay}}
    \hfil
    \subfloat[]{\includegraphics[width=0.59\linewidth]{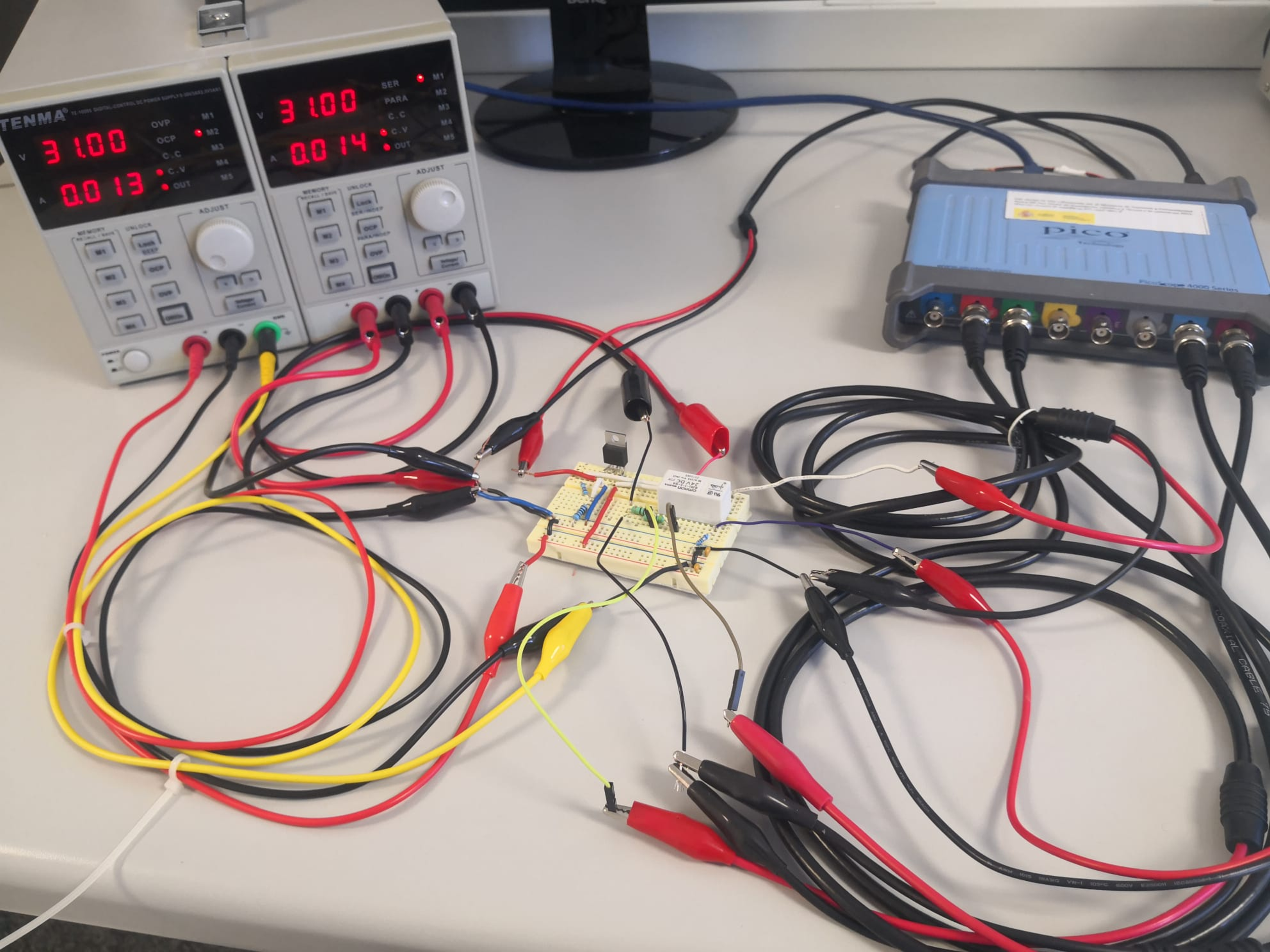}%
    \label{fig:setup}}
    \\
    \subfloat[]{\includegraphics[width=0.75\linewidth]{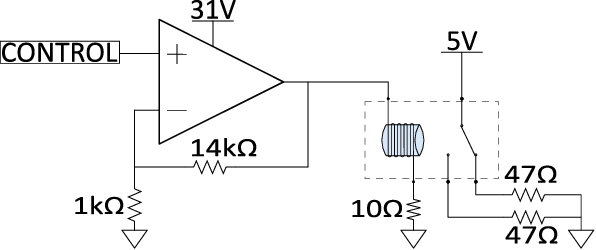}%
    \label{fig:circuit}}
    \caption{Experimental setup. (a) Tested electromagnetic relay without its enclosure. (b) Picture of the experimental setup. (c) Schematic diagram of the oscilloscope-relay link circuit.}
    \label{fig:testbech}
\vspace{\floatsep}
    \centering
    	\includegraphics[]{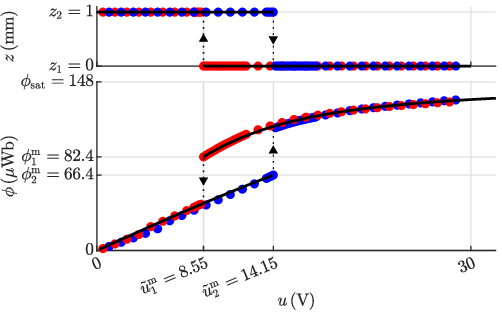}
	\caption{Experimental hysteretic switching loop. Given the symmetrical behavior, only positive voltage signals have been applied, so only the first quadrant is represented. In blue dots, trials where the initial position was $z_\mathrm{2}$. In red, trials with initial position $z_\mathrm{1}$. In black lines, model fitting.}
	\label{fig:results}
\end{figure}

The device used in the tests is a commercial single-pole double-throw (SPDT) electromagnetic relay (see Fig.~{\ref{fig:relay}}).  The experimental setup (see Fig.~{\ref{fig:setup}}) consists of a personal computer, a USB oscilloscope with a signal generator (PicoScope 4824), and an oscilloscope-relay link circuit. The personal computer, together with the oscilloscope, controls the applied signal and records the actual voltage, the relay excitation current and the status (open or closed) of the electrical contacts. The oscilloscope-relay link circuit (see Fig.~{\ref{fig:circuit}}) consists of a non-inverting amplifier stage, a current shunt, and a resistive circuit to monitor the contacts status.

To build experimentally a plot such as Fig.~{\ref{fig:hysteresis_hybrid_sat}}, the position and magnetic flux of the relay were recorded in steady state for different supply voltages. For this purpose, a series of step-type voltage signals were applied, ensuring sufficient duration to reach steady state. The magnetic flux was then calculated using {\eqref{eq:electrical}} based on the voltage and current measurements, while the armature position was determined by the state of the electrical contacts. The results, consisting of a series of triplets $(u^\mathrm{m}_{[i]},\,\phi^\mathrm{m}_{[i]},\,z^\mathrm{m}_{[i]})$, where the superscript `$\mathrm m$' denotes `measurement', are shown in Fig.~{\ref{fig:results}}. Note that the experimentally measured values of $\phi_1$, $\phi_2$, $\tilde u_1$, and $\tilde u_2$ are explicitly indicated in the axes. The results confirm the hysteretic behavior predicted by theory and clearly show the significant influence of magnetic saturation on this relay.

The close agreement between experiments and theory confirms the potential of the analysis for parameter estimation, especially since direct measurement of most of the parameters is difficult in this class of devices due to their encapsulation within a protective housing.
The goal is to estimate the value of the parameters involved in the static regime, i.e. $k_\mathrm{s}$, $z_\mathrm{s}$, $\mathcal{R}_{c0}$, $\mathcal{R}_{g0}$, $k_\mathcal{R}$, $\phi_\mathrm{sat}$, $N$, and $R$. In this particular device, the resistance, $R$, and the number of turns of the coil, $N$, are known, so only the rest of the parameters need to be estimated. This can be achieved by minimizing the following cost function:
\begin{align}
    J = &\ \sum_i \left(u^\mathrm{m}_{[i]} - \dfrac{R\,\phi^\mathrm{m}_{[i]}}{N}\left(\frac{{\mathcal{R}}_{c0}}{1-\dfrac{|\phi^\mathrm{m}_{[i]}|}{\phi_\mathrm{sat}}} + \mathcal{R}_{g0} + k_\mathcal{R} \, z^\mathrm{m}_{[i]}\right) \right)^2 \nonumber\\
    & + \alpha_\phi\big[ (\phi_1 - \phi_1^\mathrm{m} )^2 +(\phi_2 - \phi_2^\mathrm{m}  )^2\big]\nonumber\\
    & + \alpha_u\big[ (\tilde u_1 - \tilde u_1^\mathrm{m} )^2 +(\tilde u_2 - \tilde u_2^\mathrm{m} )^2 \big].
\end{align}
In this expression, the first term drives the solution toward satisfying {\eqref{eq:f3_stability}} in steady state, while the second and third terms, weighted by coefficients $\alpha_\phi$ and $\alpha_u$, aim to minimize the difference between the flux and voltage values at switching and their theoretical counterparts, given by {\eqref{eq:phi1}}, {\eqref{eq:phi2}}, {\eqref{eq:tilde_u1}}, and {\eqref{eq:tilde_u2}}. The fitting of the model to the data, also shown in Fig.~{\ref{fig:results}} in continuous black lines, proves that the optimization process was satisfactory and that the bistable behavior of the system is correctly captured by the analytical expressions previously obtained. The estimated parameter values are shown in Table~{\ref{tab:param_estimated}}.

\begin{table}[t]
	\renewcommand{\arraystretch}{1.3}
	\caption{Estimated parameter values for the electromagnetic relay.}
	\centering
    \label{tab:param_estimated}
	\begin{tabular}{ccccc}
		\cmidrule{1-2} \cmidrule{4-5}
		Parameter & Value &  &
		Parameter & Value \\ 
		\cmidrule{1-2} \cmidrule{4-5}
		$\mathcal{R}_{c0}$ & { $1.15\!\cdot\!10^{4}$ H$^{-1}$} &&
        $\phi_\mathrm{sat}$ & { $148$ $\upmu$Wb} \\
        $\mathcal{R}_{g0}$ & { $4.49\!\cdot\!10^{4}$ H$^{-1}$} &&
        $k_\mathrm{s}$ & {$94.65$ N/m} \\
        $k_\mathcal{R}$ & { $7.95\!\cdot\!10^{7}$ H$^{-1}$/m} && 
        $z_\mathrm{s}$ & {$2.85$ mm}\\
		\cmidrule{1-2} \cmidrule{4-5} 
	\end{tabular}
\end{table}

\section{Conclusions}

In this paper, we have presented a comprehensive stability analysis for electromechanical switching devices, a category encompassing electromagnetic relays and solenoid actuators. One of the key objectives was to derive expressions for the voltage and magnetic flux at which these devices switch. To do this, we have leveraged two variants of a state-of-the-art hybrid dynamical model, including magnetic saturation in one of them for enhanced accuracy at the cost of greater complexity. Both models assume a one-to-one relationship between current and magnetic flux. This simplification yields valuable analytical expressions for initial understanding, rapid analysis, and design optimization. For applications involving high-coercivity materials or high-frequency scenarios, a more comprehensive analysis incorporating magnetic hysteresis may be beneficial, although this would probably imply numerical procedures.

Experimental tests on a real device have confirmed the usefulness of the derived expressions in predicting the hysteretic relation between supply voltage, magnetic flux, and position. As a practical application, we have described and presented results of a parameter estimation method using the obtained measurements. It is important to note that parameter estimation procedures usually rely on precise measurements and require computationally intensive numerical integration of dynamic equations. By contrast, the analytical expressions in this paper are static, eliminating the need for integration and expensive sensors. The presented method is thus clearly advantageous in this regard. As a limitation, it can only be used to estimate parameters that affect the steady state of the device.

Future research should explore incorporating additional nonlinearities, such as dry friction, flux fringing, or temperature dependence. The presented stability analysis could also be readily adapted and applied to other types of electromechanical actuators. Moreover, the derived analytical expressions offer a valuable tool for optimizing the design of novel actuators.



\end{document}